\documentclass[12pt]{iopart}

\usepackage{graphicx}

\usepackage[x11names]{xcolor}

\newcommand{\domark}{%
  \vbox to 0pt{
    \kern-\dp\strutbox
    \smash{\llap{\color{red!90!black}\#\kern0.5em}}
    \vss
  }%
}

\begin{document}

\title{Highly Confined In-plane Exciton-Polaritons in Monolayer Semiconductors}

\author{Itai Epstein}
\address{ICFO-Institut de Ciencies Fotoniques, The Barcelona Institute of Science and Technology, 08860 Castelldefels (Barcelona), Spain}

\author{Andr\'{e} J. Chaves}
\address{Grupo de Materiais Semicondutores e Nanotecnologia and Departamento de F\'{i}sica, Instituto Tecnol\'{o}gico de Aeron\'{a}utica, DCTA, 12228-900 S\~{a}o Jos\'{e} dos Campos, Brazil}

\author{Daniel A. Rhodes}
\address{Department of Mechanical Engineering, Columbia University, New York, NY 10027} 

\author{Bettina Frank}
\address{4th Physics Institute and Research Center SCoPE, University of Stuttgart, 70569 Stuttgart, Germany} 

\author{Kenji Watanabe}
\address{National Institute for Materials Science, 1-1 Namiki, Tsukuba 305-0044, Japan} 

\author{Takashi Taniguchi}
\address{National Institute for Materials Science, 1-1 Namiki, Tsukuba 305-0044, Japan} 

\author{Harald Giessen}
\address{4th Physics Institute and Research Center SCoPE, University of Stuttgart, 70569 Stuttgart, Germany}

\author{James C. Hone}
\address{Department of Mechanical Engineering, Columbia University, New York, NY 10027}

\author{Nuno M. R. Peres}
\address{Centro de F\'{i}sica and Departamento de F\'{i}sica and QuantaLab, Universidade do Minho, P-4710-057 Braga, Portugal}
\address{International Iberian Nanotechnology Laboratory (INL), Av. Mestre José Veiga, 4715-330 Braga, Portugal}

\author{Frank H. L. Koppens}
\address{ICFO-Institut de Ciencies Fotoniques, The Barcelona Institute of Science and Technology, 08860 Castelldefels (Barcelona), Spain}
\address{ICREA – Instituci\'{o} Catalana de Recerca i Estudis Avan\c{c}ats, Barcelona, Spain}

\begin{abstract}
\textbf{2D materials support unique excitations of quasi-particles that consist of a material excitation and photons called polaritons. Especially interesting are in-plane propagating polaritons which can be confined to a single monolayer and carry large momentum. In this work, we report the existence of a new type of in-plane propagating polariton, supported on monolayer transition-metal-dicalcogonide (TMD) in the visible spectrum, which has not yet been observed. This 2D in-plane exciton-polariton (2DEP) is described by the coupling of an electromagnetic light field with the collective oscillations of the excitons supported by monolayer TMDs. We expose the specific experimental conditions required for the excitation of the 2DEP and show that these can be created if the TMD is encapsulated with hexagonal-boron-nitride (hBN) and cooled to cryogenic temperatures. In addition, we compare the properties of the 2DEPs with those of surface-plasmons-polaritons (SPPs) at the same spectral range, and find that the 2DEP exhibit over two orders-of-magnitude larger wavelength confinement. Finally, we propose two configurations for the possible experimental observation of 2DEPs.}
\end{abstract}

%
%
%
%

Polaritons in 2D materials have attracted great interest in the last few years, owing to their remarkable properties and to the fact that they are supported by atomically thin monolayers \cite{Basov2016,Low2017}. Different types of polaritons with different properties have been observed on a variety of 2D materials. Monolayer graphene, for example, supports extremely confined and low loss in-plane propagating polaritons in the mid-infrared (MIR) and terahertz (THz) spectral range, named graphene-plasmons (GPs) \cite{Wunsch2006,Hwang2007,Jablan_PRB_2009}. Another example is hBN, which is an anisotropic 2D material that supports large momentum, in-plane propagating hyperbolic phonon-polaritons, in its MIR Reststrahlen bands \cite{Dai2014,Caldwell2014,Dai2015}.\\

In the visible spectrum, monolayer TMDs, such as MoS$_2$, MoSe$_2$, WS$_2$ and WSe$_2$, are known to support robust excitons with large binding energies, which interact strongly with light \cite{Mueller2018,Wang2018a,Epstein2019}. When the TMD is placed in an optical cavity, these excitons can couple with cavity photons and form an out-of-plane propagating exciton-polariton \cite{Liu2014}, similarly to excitons in quantum-wells.\\

The uniqueness of polaritons arises from the fact that the new formed quasi-particle inherits both the properties of the photon and those of the particle (i.e. plasmon/phonon/exciton etc), and is thus half light and half matter in its nature. This fact enables to control one of these properties by manipulating the other. For example, it allows to probe quantum effects via classical light \cite{Lundeberg2017}, to control particle propagation by manipulating the photon wave properties \cite{Epstein2014,Epstein2014a,Epstein2016,Barachati2018}, probe bosonic condensation with exciton-polaritons \cite{Fogler2014,Kogar2017,Wang2019}, and so on. \\

The type of polariton we focus on in this study is commonly associated with transverse-magnetic (TM) surface-polaritons, residing at the interface between two bulk materials. The condition for the existence of such a mode at the interface is that one of the materials should exhibit a permittivity whose real part is negative, while the other is usually taken as a dielectric. The negative real part of the permittivity, relates the existence of charges in the material, which are polarized by the impinging electromagnetic field. For example, electron oscillations give rise to the surface-plasmon-polariton (SPP) in metal surfaces \cite{Maier2007}, or lattice oscillation in the form of optical phonons that give rise to the surface-phonon-polariton (SPhP) in polar dielectrics \cite{Caldwell2015}. All of these exhibit a negative real part of the permittivity in a certain spectral range, and the response can be directly translated from interfaces of bulk materials to bulky slabs and to monolayers (i.e. very thin slabs) \cite{Basov2016,Low2017}. Thus the fundamental question that requires answering in the case of monolayer TMDs is whether a negative real part permittivity can be achieved in these materials, and if so, what would be the properties of the formed polariton under these conditions? \\

\begin{figure*}[ht!] 
  \centering
  \includegraphics[scale=0.3]{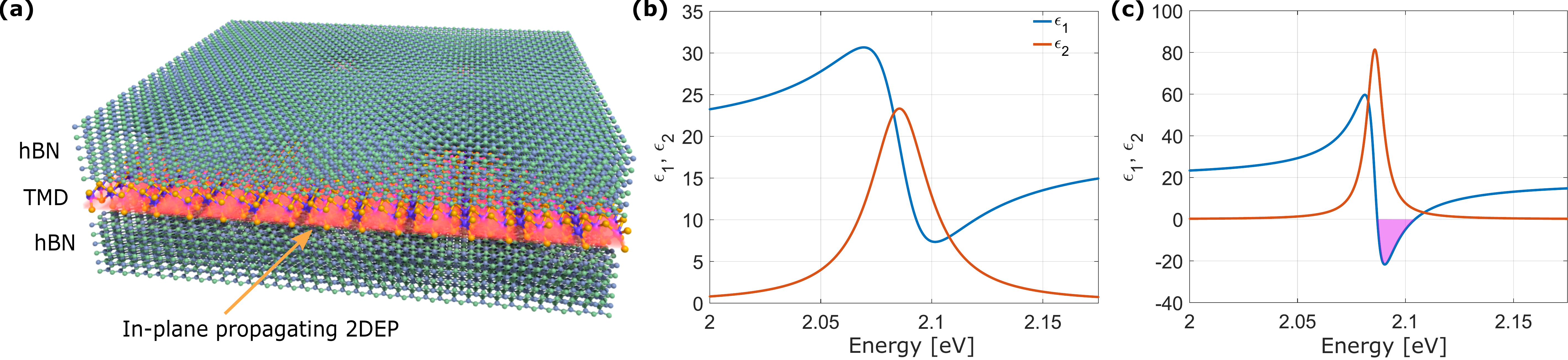} 
   \caption{Concept and conditions for existence of an in-plane 2DEP mode. (a) An hBN/monolayer TMD/hBN heterosturcute supporting an in-plane prpogating 2DEP mode, which is confined vertically to the monolayer. Real and imaginary parts of the TMD susceptibility, $\epsilon_1$ and $\epsilon_2$, respectively, showing $\epsilon_1$ with, (b) all-positive values for large exciton linewidth $\gamma_{\mathrm{T}}$, and (c) negative values that can be obtained for smaller $\gamma_{\mathrm{T}}$ (pink shaded area). 
	}
	\label{fig:figure1}
\end{figure*}

In this work, we discuss the existence of a new type of exciton-polariton in monolayer TMDs, the 2DEP, which propagates in-plane and is confined to the monolayer in the out-of-plane direction (Fig.~\ref{fig:figure1}a). It belongs to a specific family of polaritons, in which the oscillations of the impinging TM electromagnetic light field couples with collective oscillations of the TMD excitons, and indeed exhibits the known signature of negative real part of its permittivity. We show that owing to the resonant nature of the excitonic optical response in TMDs, the formation of the 2DEP and its detection strongly depend on achieving extremely narrow excitonic linewidths. Furthermore, we demonstrate that the conditions for the formation of the 2DEP are experimentally attainable by encapsulating the monolayer TMD with hBN, and cooling down to cryogenic temperatures. In addition, we analyze the 2DEP properties under these conditions and compare them to those of SPPs at the same spectral range. We find that the 2DEP exhibit over two orders-of-magnitude larger wavelength confinement, which is accompanied with about four time larger propagation damping. Finally, we propose the usage of TMD nanoribbons and metal gratings, as two possible methods for the experimental observation of 2DEPs.  \\


The optical response of TMD excitons in usually characterized by a Lorentzian complex susceptibility, taking the form \cite{Scuri2018}:

\begin{equation}
\chi(\omega)=\chi_{\mathrm{bg}}-\frac{\mathrm{c}}{\omega_{\mathrm{0}}\mathrm{d}}\frac{\gamma_{\mathrm{r,0}}}{\omega-\omega_{\mathrm{0}}+i\left(\frac{\gamma_{\mathrm{nr}}}{2}+\gamma_{\mathrm{d}}\right)},
\label{eq:1}
\end{equation}

where $\chi_{\mathrm{bg}}$ is the background susceptibility, c is the speed of light, $\omega_{\mathrm{0}}$ is the exciton energy, d is the monolayer thickness, and $\gamma_{\mathrm{r,0}}$, $\gamma_{\mathrm{nr}}$, $\gamma_{\mathrm{d}}$ are the radiative, non-radiative and pure dephasing decay rates, respectively. $\gamma_{\mathrm{nr}}$ and $\gamma_{\mathrm{d}}$ correspond to different physical loss channels, which affect the optical response of the TMD in different manners. The approach previously introduced in \cite{Epstein2019} allows for their separate modeling (and thus extraction from experimental data), with the total linewidth for a suspended TMD being defined as $\gamma_{\mathrm{T}}=\gamma_{\mathrm{r,0}}$+$\gamma_{\mathrm{nr}}$+2$\gamma_{\mathrm{d}}$ .\\

The frequency dependent permittivity obtained from Eq.~\ref{eq:1}, i.e. $1+\chi(\omega)$, is presented in Fig.~\ref{fig:figure1}b, indicating the behavior of both the real and imaginary parts of the TMD, $\epsilon_1$ and $\epsilon_2$, respectively, at the spectral region of the exciton resonance $\omega_0$. It can be seen in Fig.~\ref{fig:figure1}b that $\epsilon_1$ is positive over the complete spectral range, which indeed corresponds to what have been previously measured for non-encapsulated monolayer TMDs at room temperature \cite{Li2014}. \\

\begin{figure*}[ht!] 
  \centering
  \includegraphics[scale=0.45]{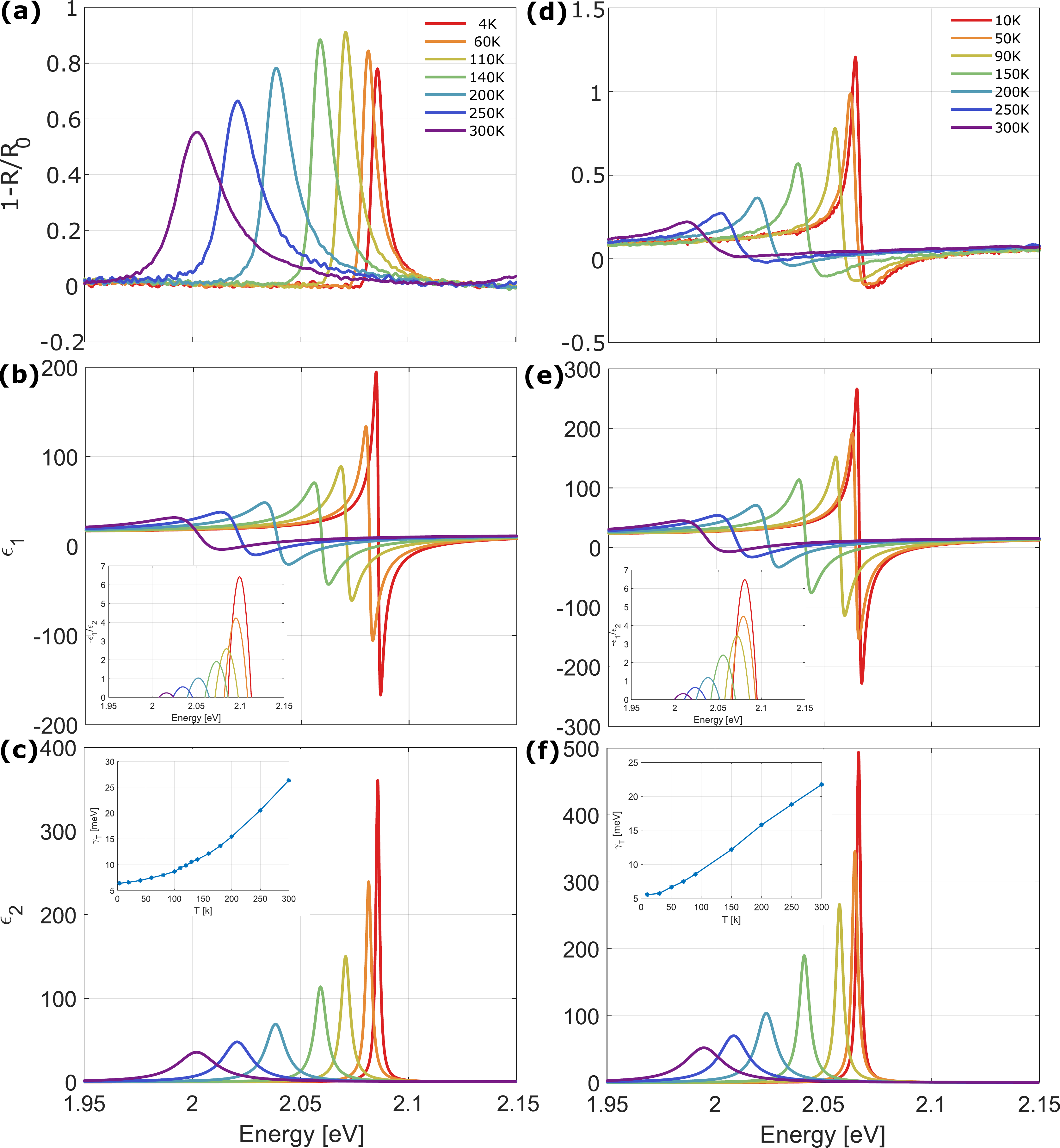} 
   \caption{Temperature dependent experimental measurements of the reflection contrast and extracted complex permittivity for two different samples. (a) Reflection contrast measured for sample U1, with 15 nm hBN/monolayer WS$_2$/30 nm hBN/gold mirror structure, and (d) for sample U2, with 25 nm hBN/monolayer WS$_2$/45 nm hBN/Sapphire substrate. (b),(c) and (e),(f) extracted temperature dependent $\epsilon_1$ and $\epsilon_2$, for samples U1 and U2, respectively. The insets in (b) and (e) correspond to the ratio $-\epsilon_1/\epsilon_2$, signifying the optimal area of confinement and loss.The insets in (c) and (f) correspond to the temperature dependent linewidth, $\gamma_{\mathrm{T}}$, measured for samples U1 and U2, respectively. See methods section for the extracted parameters. 
	}
	\label{fig:figure2}
\end{figure*}

Whether $\epsilon_1$ takes positive or negative values depends on two main attributes. The first is the background susceptibility $\chi_{\mathrm{bg}}$, which is an intrinsic property that depends on all higher energy optical resonances of the material. It basically shifts $\epsilon_1$ to higher positive values, as it corresponds to the value of $\epsilon_1$ in the absence of the excitonic resonance. The second is the total linewidth of the exciton resonance, $\gamma_{\mathrm{T}}$, which affects the amplitude of the optical response through $\gamma_{\mathrm{r,0}}$, $\gamma_{\mathrm{nr}}$, and $\gamma_{\mathrm{d}}$. For small enough $\gamma_{\mathrm{T}}$, the amplitude of both $\epsilon_1$ and $\epsilon_2$ increases significantly, allowing $\epsilon_1$ to attain negative values (Fig.~\ref{fig:figure1}c). \\

The value of $\gamma_{\mathrm{T}}$ is determined by the homogenous linewidth of the exciton, which depends on the TMD quality \cite{Rhodes2019}, and on the interaction with other quasi-particles, such as phonons and trions, and the charge carrier density in the TMD \cite{Mak2013a,Selig2016,Ajayi2017,Cadiz2017}. Therefore, $\gamma_{\mathrm{T}}$ can be controlled via temperature, affecting the phonon contribution to $\gamma_{\mathrm{nr}}$, and electrical doping of the TMD, which modifies the formation of trions and electron-electron interactions. It has been shown that combining high quality TMDs with encapsulation in hBN (Fig.~\ref{fig:figure1}a) and cooling down to cryogenic temperatures indeed yields narrow excitonic linewidths \cite{Ajayi2017,Cadiz2017,Epstein2019}. However, the complex frequency dependent optical properties of TMDs at these conditions have not been studied.\\ 

\begin{figure*}[ht!] 
  \centering
  \includegraphics[scale=0.35]{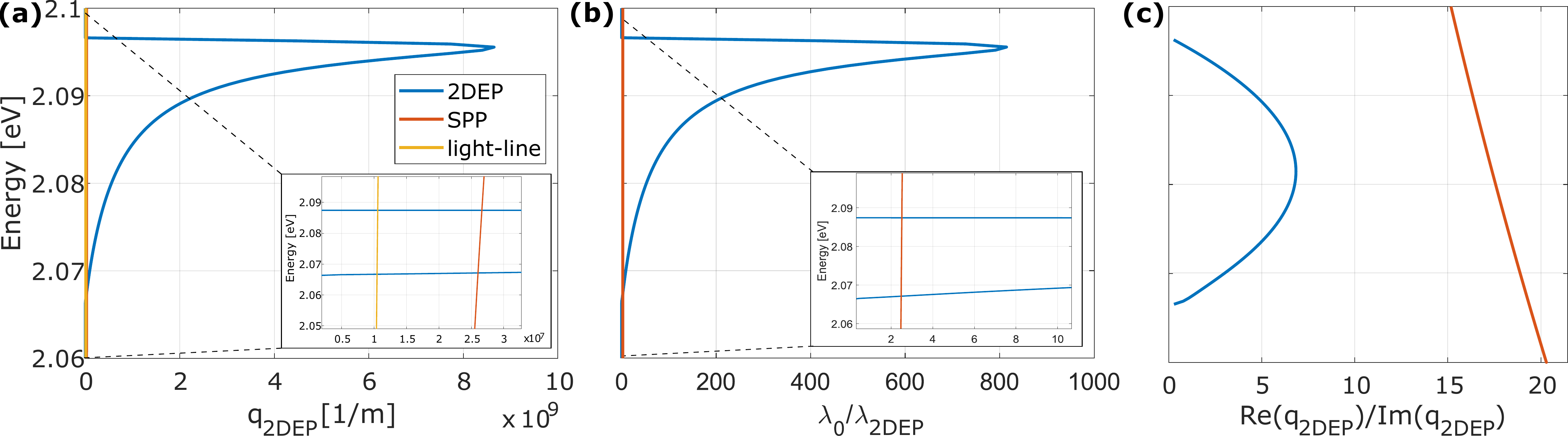} 
   \caption{Polaritonic characteristics of 2DEPs and their comparison to SPPs. (a) Calculated dispersion relation of 2DEP (blue), SPP (red) and the light-line (yellow), showing very large momentum, $\mathrm{q}_{\mathrm{2DEP}}$, carried by the 2DEP mode compared to the SPP and the light-line. The inset shows the zoomed-in, low momentum area, where the SPP and the light-line dispersion relation resides. (b) Extracted confinement factor, $\lambda_0/\lambda_{\mathrm{2DEP}}$, showing over two orders-of-magnitude larger confinement of 2DEP compared to the SPP (inset). (c) Extracted Re($\mathrm{q}_{\mathrm{2DEP}}$)/Im($\mathrm{q}_{\mathrm{2DEP}}$), showing that the 2DEP has on average 4 times larger propagation damping, compared to the SPP. In these calculations, the extracted permittivity from sample U2 wass used for the hBN encapsulated WS$_2$, and for an SPP at the interface between hBN and Au \cite{Maier2007}.
	}
	\label{fig:figure3}
\end{figure*}

Fig.~\ref{fig:figure2}a,d shows the measured temperature dependent reflection extinction, $1-\mathrm{R/R_0}$, for two different samples of monolayer WS$_2$ encapsulated by hBN. The different spectral response observed in Fig.~\ref{fig:figure2}a,d stems from the different structures of the samples. Sample U1 (Fig.~\ref{fig:figure2}a-c) consists of an hBN/TMD/hBN heterostructure placed on a single crystalline, atomically flat gold mirror \cite{Podbiel2019}, with hBN thicknesses chosen to maximize the excitonic absorption \cite{Epstein2019}, resulting in a Voigt-like line-shape. For sample U2 (Fig.~\ref{fig:figure2}d-f), a heterostructure with arbitrary hBN thicknesses was placed on a sapphire substrate, and the resulting Fano-like line-shape stems from the interferences in the multi-layered structure \cite{Scuri2018}.\\

For both cases, we use a specialized transfer-matrix-method to calculate the reflectance coefficient from the different heterostructures, with the TMD excitonic response being described by Eq.~\ref{eq:1}. To address pure dephasing $\gamma_d$ independently from $\gamma_{nr}$, we account for processes that change the phase of the exciton's wavefunction, without affecting the total number of excitons \cite{Epstein2019}. From the experimental data measured in Fig.~\ref{fig:figure2}a,d we extract $\gamma_{r,0}$,$\gamma_{nr}$, $\gamma_d$, and $\chi_{\mathrm{bg}}$ using the same approach as in \cite{Epstein2019} (see methods), and these are then used in Eq.~\ref{eq:1} in order to obtain $\chi(\omega)$ and $\epsilon(\omega)$. \\

The resulting frequency-dependent real and imaginary parts of the permittivity, $\epsilon_1$ and $\epsilon_2$, are presented in Fig.~\ref{fig:figure2}b,c,e,f. For both samples, it can be seen that already at room temperature $\epsilon_1$ is slightly negative. This stems from the high quality TMDs used in these samples (see methods), as well as their encapsulation with hBN, which result in narrow $\gamma_{\mathrm{T}}$ (insets of Fig.~\ref{fig:figure2}c,f), in contrast to an all-positive $\epsilon_1$ previously measured for non-encapsulated, commercially obtained TMDs \cite{Li2014}. It can also be seen that for both samples, decreasing the temperature leads to a reduction of $\gamma_{\mathrm{T}}$, and at the same time $\epsilon_1$ gradually attains increasing negative values at the high energy region of the resonance.\\

This observed behavior implies that monolayer TMDs can indeed support propagating 2DEPs under these conditions. It can further be seen in Fig.~\ref{fig:figure2}b,c,e,f, that the majority of the spectral region where $\epsilon_1$ attains negative values corresponds to the spectral region where $\epsilon_2$ is large, which translates to increased losses due to absorption. This observation implies that the losses experienced by the excited 2DEPs in the system will depend strongly on $\gamma_{\mathrm{T}}$. The optimal case for such a resonant system is that $\gamma_{\mathrm{T}}$ is small enough to provide a spectral region where $\epsilon_1$ is still negative and the value of $\epsilon_2$ is very low, exhibited by the calculated ratio $-\epsilon_1/\epsilon_2$ in the insets of Fig.~\ref{fig:figure2}b,e. For this reason as well, it is crucial to obtain as narrow as possible exciton linewidths.\\

Next we analyze the 2DEP properties that are obtained from the extracted permittivity in Fig.~\ref{fig:figure2}, for the lowest temperature of sample U2. We present in Fig.~\ref{fig:figure3}a the numerical solution for the dispersion relation for an hBN encapsulated WS$_2$ monolayer, with the permittivity obtained from sample U2. The 2DEP mode can be clearly seen, carrying larger momentum than the free-space photon, i.e., the light-line (yellow line), as expected from the dispersion relation of this type of polaritons. It can further be seen that the momentum carried by the 2DEP is also significantly larger than that of a SPP in the same spectral range (red line). \\

As SPPs are known for their abilities to carry high momentum and confine and enhance the optical field \cite{Maier2007}, it is worthwhile to compare their properties to those attained by the 2DEP, especially since they share the same spectral range. Fig.~\ref{fig:figure3}b,c show the extracted confinement factor, $\lambda_{\mathrm{mode}}/\lambda_{\mathrm{0}}$, and propagation damping factor, Re($\mathrm{q}_{\mathrm{mode}}$)/Im($\mathrm{q}_{\mathrm{mode}}$), for the two modes, where the mode corresponds to either the 2DEP or SPP, and $\lambda_{\mathrm{0}}$ is the free-space wavelength. It can be seen in Fig.~\ref{fig:figure3}b that the 2DEP confinement factor is over two orders-of-magnitude larger, which results in a 2DEP group velocity that is $2.5\cdot\mathrm{10^{6}}$ times smaller than the speed of light. This group velocity is four orders-of-magnitude slower than what has been observed for waveguided exciton-polaritons in TMDs \cite{Mrejen2019}. However, this increased confinement factor is associated with an increase in the propagation damping of about a factor of four compared to SPPs (Fig.~\ref{fig:figure3}c). This corresponds to an average of $\sim$15 nm propagation length, signifying again the importance of achieving excitonic linewidths as narrow as possible. The latter can be obtained by using other TMDs, such as MoSe$_2$ and WSe$_2$, which were shown to reach exciton linewidth close to $1$ meV \cite{Ajayi2017,Cadiz2017}, compared to the best achieved here of $5.5$ meV for WS$_2$ in sample U2. \\

\begin{figure*}[ht!] 
  \centering
  \includegraphics[scale=0.4]{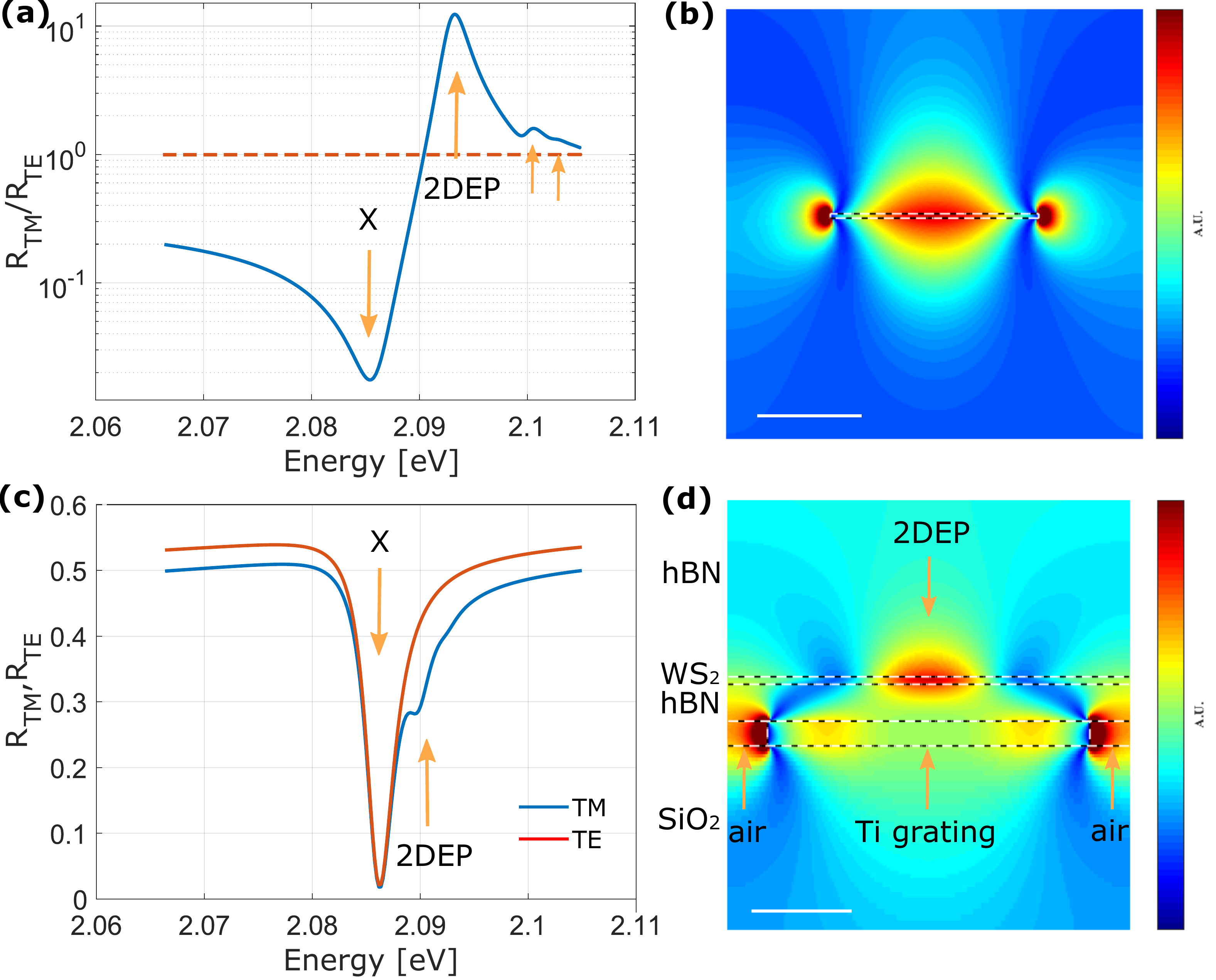} 
   \caption{Experimental approaches for the observation of 2DEPs. (a) Simulated normalized TM/TE reflection spectrum obtained for 40 nm monolayer WS$_2$ nanoribbons, showing the exciton location (marked as X), and several resonances of the 2DEP supported by the nanoribbons (marked as 2DEP and orange arrows). The dashed red line corresponds to the case of an unpatterned monolayer. (b) Electric field distribution, $|\mathrm{E}_{\mathrm{x}}|$,  of the first order resonance of the WS$_2$ nanoribbon, marked by the orange arrow in (a), showing the standing wave pattern formed by the 2DEP. The WS$_2$ nanoribbon is marked by the dashed white lines. Scale bar is 20 nm. (c) The TM (blue curve) and TE (red curve) reflection spectrum obtained from an hBN encapsulated WS$_2$, placed over a 2 nm thick Ti grating, with 40 nm period and 80$\%$ duty-cycle, showing the exciton location (marked as X), and the additional 2DEP resonance obtained only for TM polarization, marked by the orange arrow. (d) Electric field distribution, $|\mathrm{E}_{\mathrm{x}}|$, obtained at resonance, showing the formation of the 2DEP mode. The WS$_2$ and Ti grating are marked by the dashed white lines. Scale bar is 10 nm.
	}
	\label{fig:figure4}
\end{figure*}

Next we propose two possible methods for experimentally observing the 2DEP mode. First, we pattern the monolayer TMD into nanoribbons, which support standing waves formed by the propagating 2DEPs. Fig.~\ref{fig:figure4}a depicts the simulated normalized TM/TE refection spectra obtained for $40$ nm nanoribbons of WS$_2$, exhibiting several resonances on the high energy side of the exciton resonance, corresponding to the spectral range where $\epsilon_1$ is negative. The dashed line corresponds to the case of an unpatterned monolayer. An examination of the electric field distribution of the first order resonance (marked by the orange arrow) is presented in Fig.~\ref{fig:figure4}b, displaying the formed standing wave pattern. This corroborates the excitation of the 2DEP mode, similar to those observed for graphene-plasmons on graphene nanoribbons \cite{Brar2014}. The additional resonances appearing in Fig.~\ref{fig:figure4}a correspond to higher order 2DEPs.\\

Patterning the TMD to nanoribbons will also affect its optical response, as it usually creates rough edges, which lowers the quality of the TMD and therefore broadens the linewidth. Owing to this, we propose a second approach which does not require patterning and is based on transferring the hBN/TMD/hBN heterostructure on top of a thin metallic grating. The bottom hBN thickness is chosen to be $3$ nm in this case, in order to ensure efficient excitation. Fig.~\ref{fig:figure4}c shows the obtained TM and TE reflection spectra, exhibiting an additional resonance (marked by the orange arrow), which only exists in TM polarization, as expected from this type of polaritons. \\

To corroborate its nature, we present its electric field distribution in Fig.~\ref{fig:figure4}d, showing a more complex pattern. In this case, 2DEPs are excited both above the metallic part and above the non-metallic part, with different properties \cite{Iranzo2018}. The reason is that image charges in the metallic part screen the charge carriers in the TMD, altering the properties of the mode and its dispersion relation, similar to graphene-plasmons on a graphene sheet placed close to a metallic grating \cite{Iranzo2018,Epstein2020}. Since screened 2DEPs are excited above the metallic part and unscreened 2DEPs above the non-metallic part, their interaction is more complex and requires a more in depth analysis. However, as these are distinct properties of propagating polaritons they corroborate the excitation of 2DEPs in this system. \\

In conclusion, we have reported here the existence of a new type of in-plane propagating exciton-polaritons on monolayers TMDs, which have not been previously observed. We have identified the conditions that are required for its excitation and also provided experimental evidence that these can be achieved. In addition, we have analyzed its polaritonic properties and compared them with those of SPPs, finding over two orders-of-magnitude larger 2DEP confinement factors. Finally, we proposed and numerically demonstrated two possible approaches for the experimental observation of 2DEPs. The results and analysis provided here paves they way for the experimental observation of the 2DEP. With the use of other TMDs supporting narrower linewidths, lower losses might be obtained, improving further the characteristics of the 2DEP presented here.   \\

\section*{\textbf{Methods}}

	\textbf{Optical setup}
Temperature dependent measurements were done in an Attodry800 cryostat, and spectral detection with an Andor spectrometer, with a white light source. The light was focused using a Nikon objective with $\mathrm{NA}=0.6$.\\

\textbf{Synthesis}
WS$_2$ was grown using chemical vapor transport with a WCl$_6$ precursor in excess sulfur. W, 99.999 \%, and sulfur (99.9995 \%) were first loaded into a quartz ampoule in a 1:2, W:S, ratio with an additional excess of 24 mg/cm$^3$ of sulfur. 60 mg of WCl$_6$ was added into the quartz ampoule inside of a nitrogen glovebox, the ampoule was then removed from the glovebox and quickly pumped down to 5 x 10$^{-5}$ Torr to avoid deterioration of the WCl$_6$. The ampoule was then sealed under vacuum.  The reagents were subsequently heated to 1000 $^{\circ}$C within 24 h and with a 100 $^{\circ}$C temperature gradient between the hot and cold zone. The ampoule was then allowed to dwell at this temperature for 2 weeks before being cooled to 400 $^{\circ}$C over an additional two weeks. The as harvested crystals were then rinsed in acetone and isopropanol and dried in air before being used.   \\

\textbf{Device Fabrication}
Monolayer flakes of WS$_2$ were mechanically exfoliated from the aforementioned single crystals on to 285 nm SiO$_2$ chip. hBN was exfoliated and picked up using a dry transfer, PDMS/polypropylene carbonate (PPC) stamp method mounted on to a glass slide. The final stack is then transferred on to a the gold mirror/Sapphire at 125 $^{\circ}$C and rinsed in acetone.   \\

\textbf{Extracted data for Sample U1}
We find that $\chi_{\mathrm{bg}}=12$ and $\gamma_{r,0}=2.43$ meV. For the refractive index of the substrate $\sqrt{\epsilon_\mathrm{hBN}}=2.2$ have been used, and for gold we use the same model used in \cite{Epstein2019}.

\section*{\textbf{Acknowledgments}}
I.E. thanks Dr. Fabien Vialla. J.H. and D.R. acknowledge the funding support by the NSF MRSEC program through Columbia in the Center for Precision Assembly of Superstratic and Superatomic Solids (DMR-1420634). H.G. and B.F. acknowledge support from ERC advanced grant COMPLEXPLAS. F.H.L.K. acknowledges financial support from the Government of Catalonia trough the SGR grant, and from the Spanish Ministry of Economy and Competitiveness, through the “Severo Ochoa” Programme for Centres of Excellence in R\&D (SEV-2015-0522), support by Fundacio Cellex Barcelona, Generalitat de Catalunya through the CERCA program,  and the Mineco grants Plan Nacional (FIS2016-81044-P) and the Agency for Management of University and Research Grants (AGAUR) 2017 SGR 1656.  Furthermore, the research leading to these results has received funding from the European Union Seventh Framework Programme under grant agreement no.785219 (Core2) and no. 881603 (Core3) Graphene Flagship. This work was supported by the ERC TOPONANOP under grant agreement no. 726001.

\section*{\textbf{References}}
\bibliography{2DEP}

\providecommand{\newblock}{}
\begin{thebibliography}{10}
\expandafter\ifx\csname url\endcsname\relax
  \def\url#1{{\tt #1}}\fi
\expandafter\ifx\csname urlprefix\endcsname\relax\def\urlprefix{URL }\fi
\providecommand{\eprint}[2][]{\url{#2}}

\bibitem{Basov2016}
Basov D~N, Fogler M~M and {Garc{\'{i}}a De Abajo} F~J 2016 {\em Science\/} {\bf
  354} 195 ISSN 10959203

\bibitem{Low2017}
Low T, Chaves A, Caldwell J~D, Kumar A, Fang N~X, Avouris P, Heinz T~F, Guinea
  F, Martin-Moreno L and Koppens F 2017 {\em Nature Materials\/} {\bf 16}
  182--194 ISSN 14764660 (\textit{Preprint} \eprint{1610.04548})

\bibitem{Wunsch2006}
Wunsch B, Stauber T, Sols F and Guinea F 2006 {\em New Journal of Physics\/}
  {\bf 8} ISSN 13672630 (\textit{Preprint} \eprint{0610630})

\bibitem{Hwang2007}
Hwang E~H and {Das Sarma} S 2007 {\em Phys. Rev. B\/} {\bf 75} 205418 ISSN
  1098-0121 \urlprefix\url{http://link.aps.org/doi/10.1103/PhysRevB.75.205418}

\bibitem{Jablan_PRB_2009}
Jablan M, Buljan H and Solja{\v{c}}i{\'{c}} M 2009 {\em Physical Review B\/}
  {\bf 80} 245435 ISSN 1098-0121
  \urlprefix\url{http://prb.aps.org/abstract/PRB/v80/i24/e245435}

\bibitem{Dai2014}
Dai S, Fei Z, Ma Q, Rodin A~S, Wagner M, McLeod A~S, Liu M~K, Gannett W, Regan
  W, Watanabe K, Taniguchi T, Thiemens M, Dominguez G, {Castro Neto} A~H, Zettl
  A, Keilmann F, Jarillo-Herrero P, Fogler M~M and Basov D~N 2014 {\em
  Science\/} {\bf 343} 1125--1129 ISSN 10959203

\bibitem{Caldwell2014}
Caldwell J~D, Kretinin A~V, Chen Y, Giannini V, Fogler M~M, Francescato Y,
  Ellis C~T, Tischler J~G, Woods C~R, Giles A~J, Hong M, Watanabe K, Taniguchi
  T, Maier S~A and Novoselov K~S 2014 {\em Nature Communications\/} {\bf 5}
  1--9 ISSN 20411723

\bibitem{Dai2015}
Dai S, Ma Q, Liu M~K, Andersen T, Fei Z, Goldflam M~D, Wagner M, Watanabe K,
  Taniguchi T, Thiemens M, Keilmann F, Janssen G~C, Zhu S~E, Jarillo-Herrero P,
  Fogler M~M and Basov D~N 2015 {\em Nature Nanotechnology\/} {\bf 10} 682--686
  ISSN 17483395 (\textit{Preprint} \eprint{1501.06956})

\bibitem{Mueller2018}
Mueller T and Malic E 2018 {\em npj 2D Materials and Applications\/} {\bf 2}
  1--12 ISSN 23977132

\bibitem{Wang2018a}
Wang G, Chernikov A, Glazov M~M, Heinz T~F, Marie X, Amand T and Urbaszek B
  2018 {\em Reviews of Modern Physics\/} {\bf 90} 021001 ISSN 15390756
  (\textit{Preprint} \eprint{1707.05863})

\bibitem{Epstein2019}
Epstein I, Terr{\'{e}}s B, Chaves A~J, Pusapati V~V, Rhodes D~A, Frank B,
  Zimmermann V, Qin Y, Watanabe K, Taniguchi T, Giessen H, Tongay S, Hone J~C,
  Peres N~M~R and Koppens F 2019 {\em arXiv:1908.07598\/} (\textit{Preprint}
  \eprint{1908.07598}) \urlprefix\url{http://arxiv.org/abs/1908.07598}

\bibitem{Liu2014}
Liu X, Galfsky T, Sun Z, Xia F, Lin E~C, Lee Y~H, K{\'{e}}na-Cohen S and Menon
  V~M 2014 {\em Nature Photonics\/} {\bf 9} 30--34 ISSN 17494893
  (\textit{Preprint} \eprint{1406.4826})

\bibitem{Lundeberg2017}
Lundeberg M~B, Gao Y, Asgari R, Tan C, Duppen B~V, Autore M,
  Alonso-Gonz{\'{a}}lez P, Woessner A, Watanabe K, Taniguchi T, Hillenbrand R,
  Hone J, Polini M and Koppens F~H~L 2017 {\em Science\/} {\bf 357} 187--191
  ISSN 10959203 (\textit{Preprint} \eprint{1704.05518})

\bibitem{Epstein2014}
Epstein I, Lilach Y and Arie A 2014 {\em Journal of the Optical Society of
  America B\/} {\bf 31} 1642 ISSN 0740-3224

\bibitem{Epstein2014a}
Epstein I and Arie A 2014 {\em Physical Review Letters\/} {\bf 112} 023903 ISSN
  00319007 (\textit{Preprint} \eprint{1311.0261})

\bibitem{Epstein2016}
Epstein I, Tsur Y and Arie A 2016 {\em Laser and Photonics Reviews\/} {\bf 10}
  360--381 ISSN 18638880
  \urlprefix\url{http://doi.wiley.com/10.1002/lpor.201500242}

\bibitem{Barachati2018}
Barachati F, Fieramosca A, Hafezian S, Gu J, Chakraborty B, Ballarini D,
  Martinu L, Menon V, Sanvitto D and K{\'{e}}na-Cohen S 2018 {\em Nature
  Nanotechnology\/} {\bf 13} 906--909 ISSN 17483395 (\textit{Preprint}
  \eprint{1803.04352})

\bibitem{Fogler2014}
Fogler M~M, Butov L~V and Novoselov K~S 2014 {\em Nature Communications\/} {\bf
  5} 1--5 ISSN 20411723 (\textit{Preprint} \eprint{1404.1418})

\bibitem{Kogar2017}
Kogar A, Rak M~S, Vig S, Husain A~A, Flicker F, Joe Y~I, Venema L, MacDougall
  G~J, Chiang T~C, Fradkin E, {Van Wezel} J and Abbamonte P 2017 {\em
  Science\/} {\bf 358} 1314--1317 ISSN 10959203

\bibitem{Wang2019}
Wang Z, Rhodes D~A, Watanabe K, Taniguchi T, Hone J~C, Shan J and Mak K~F 2019
  {\em Nature\/} {\bf 574} 76--80 ISSN 14764687

\bibitem{Maier2007}
Maier S~A 2007 {\em {Plasmonics: Fundamentals and applications}\/} (Springer
  US) ISBN 0387331506

\bibitem{Caldwell2015}
Caldwell J~D, Lindsay L, Giannini V, Vurgaftman I, Reinecke T~L, Maier S~A and
  Glembocki O~J 2015 {\em Nanophotonics\/} {\bf 4} 44--68 ISSN 21928606

\bibitem{Scuri2018}
Scuri G, Zhou Y, High A~A, Wild D~S, Shu C, {De Greve} K, Jauregui L~A,
  Taniguchi T, Watanabe K, Kim P, Lukin M~D and Park H 2018 {\em Physical
  Review Letters\/} {\bf 120} 037402 ISSN 10797114

\bibitem{Li2014}
Li Y, Chernikov A, Zhang X, Rigosi A, Hill H~M, {Van Der Zande} A~M, Chenet
  D~A, Shih E~M, Hone J and Heinz T~F 2014 {\em Physical Review B - Condensed
  Matter and Materials Physics\/} {\bf 90} 205422 ISSN 1550235X

\bibitem{Rhodes2019}
Rhodes D, Chae S~H, Ribeiro-Palau R and Hone J 2019 {\em Nature Materials\/}
  {\bf 18} 541--549 ISSN 14764660

\bibitem{Mak2013a}
Mak K~F, He K, Lee C, Lee G~H, Hone J, Heinz T~F and Shan J 2013 {\em Nature
  Materials\/} {\bf 12} 207--211 ISSN 1476-1122
  \urlprefix\url{http://www.nature.com/articles/nmat3505}

\bibitem{Selig2016}
Selig M, Bergh{\"{a}}user G, Raja A, Nagler P, Sch{\"{u}}ller C, Heinz T~F,
  Korn T, Chernikov A, Malic E and Knorr A 2016 {\em Nature Communications\/}
  {\bf 7} 13279 ISSN 2041-1723
  \urlprefix\url{http://www.nature.com/articles/ncomms13279}

\bibitem{Ajayi2017}
Ajayi O~A, Ardelean J~V, Shepard G~D, Wang J, Antony A, Taniguchi T, Watanabe
  K, Heinz T~F, Strauf S, Zhu X~Y and Hone J~C 2017 {\em 2D Materials\/} {\bf
  4} 1--16 ISSN 20531583 (\textit{Preprint} \eprint{1702.05857})

\bibitem{Cadiz2017}
Cadiz F, Courtade E, Robert C, Wang G, Shen Y, Cai H, Taniguchi T, Watanabe K,
  Carrere H, Lagarde D, Manca M, Amand T, Renucci P, Tongay S, Marie X and
  Urbaszek B 2017 {\em Physical Review X\/} {\bf 7} 021026 ISSN 2160-3308
  \urlprefix\url{http://link.aps.org/doi/10.1103/PhysRevX.7.021026}

\bibitem{Podbiel2019}
Podbiel D, Kahl P, Frank B, Davis T~J, Giessen H, Hoegen M~H~v and {Meyer zu
  Heringdorf} F~J 2019 {\em ACS Photonics\/} {\bf 6} 600--604 ISSN 2330-4022
  \urlprefix\url{http://pubs.acs.org/doi/10.1021/acsphotonics.8b01565}

\bibitem{Mrejen2019}
Mrejen M, Yadgarov L, Levanon A and Suchowski H 2019 {\em Science Advances\/}
  {\bf 5} eaat9618 ISSN 23752548

\bibitem{Brar2014}
Brar V~W, Jang M~S, Sherrott M, Kim S, Lopez J~J, Kim L~B, Choi M and Atwater H
  2014 {\em Nano Letters\/} {\bf 14} 3876--3880 ISSN 1530-6984
  \urlprefix\url{https://pubs.acs.org/doi/10.1021/nl501096s}

\bibitem{Iranzo2018}
Iranzo D~A, Nanot S, Dias E~J~C, Epstein I, Peng C, Efetov D~K, Lundeberg M~B,
  Parret R, Osmond J, Hong J~Y, Kong J, Englund D~R, Peres N~M~R and Koppens
  F~H~L 2018 {\em Science\/} {\bf 295} 291--295 ISSN 0036-8075
  (\textit{Preprint} \eprint{1804.01061})
  \urlprefix\url{http://arxiv.org/abs/1804.01061}

\bibitem{Epstein2020}
Epstein I, Alcaraz D, Huang Z, Pusapati V~V, Hugonin J~P, Kumar A, Deputy X,
  Khodkov T, Rappoport T~G, Peres N~M~R, Smith D~R and Koppens F~H~L 2020 {\em
  arXiv:2002.00366\/} (\textit{Preprint} \eprint{2002.00366})
  \urlprefix\url{http://arxiv.org/abs/2002.00366}

\end{thebibliography}

\end{document}